\documentclass[aps,prd,twocolumn,superscriptaddress]{revtex4-1} 
\usepackage{fullpage, amsmath, xparse, amssymb, mathtools, graphicx, braket,xcolor,verbatim}
\usepackage{natbib}
\usepackage[colorlinks=true,linkcolor=blue,citecolor=blue]{hyperref}

\usepackage{multirow}
\usepackage{hyperref}
\usepackage{cancel}

\begin{document}
\title{On the single classical field description of interacting scalar fields}
\author{Andrew Eberhardt}
\email{Corresponding author. \\ aeberhar@stanford.edu}
\affiliation{Kavli Institute for Particle Astrophysics and Cosmology, Menlo Park, 94025, California, USA}
\affiliation{Physics Department, Stanford University, Stanford, California, USA}
\affiliation{SLAC National Accelerator Laboratory}
\author{Alvaro Zamora}
\affiliation{Kavli Institute for Particle Astrophysics and Cosmology, Menlo Park, 94025, California, USA}
\affiliation{Physics Department, Stanford University, Stanford, California, USA}
\affiliation{SLAC National Accelerator Laboratory}
\author{Michael Kopp}
\affiliation{Nordita,
KTH Royal Institute of Technology and Stockholm University,
Hannes Alfv\'ens v\"ag 12, SE-106 91 Stockholm, Sweden}
\author{Tom Abel}
\affiliation{Kavli Institute for Particle Astrophysics and Cosmology, Menlo Park, 94025, California, USA}
\affiliation{Physics Department, Stanford University, Stanford, California, USA}
\affiliation{SLAC National Accelerator Laboratory}
\begin{abstract}

We test the degree to which interacting Bosonic systems can be approximated by a classical field as total occupation number is increased. This is done with our publicly available code repository, \href{https://github.com/andillio/QIBS}{QIBS}, a massively parallel solver for these systems. We use a number of toy models well studied in the literature and track when the classical field description admits quantum corrections, called the quantum breaktime. This allows us to test claims in the literature regarding the rate of convergence of these systems to the classical evolution. We test a number of initial conditions, including coherent states, number eigenstates, and field number states. We find that of these initial conditions, only number eigenstates do not converge to the classical evolution as occupation number is increased. We find that systems most similar to scalar field dark matter exhibit a logarithmic enhancement in the quantum breaktime with total occupation number. Systems with contact interactions or with field number state initial conditions, and linear dispersions, exhibit a power law enhancement. Finally, we find that the breaktime scaling depends on both model interactions and initial conditions.
\end{abstract}

\maketitle

\section{Introduction} \label{sec:Intro}

Many interesting physical systems involve a large number of interacting bosons. These include Bose Einstein condensates (BEC) \cite{Gross, Leggett2001,Anderson1995, Davis1995}, electromagnetic radiation \cite{bial1977}, and scalar field dark matter (SFDM) \cite{Hu2000, Mocz2019, guth2015, chakrabarty2021}. When occupation numbers are very large compared to unity, it is often argued that a classical, mean field theory approximation (MFT) will accurately describe field expectation values and number densities \cite{Hu2000, Mocz2019, guth2015, bial1977, Gross, Leggett2001, Hertzberg2016, Minguzzi2004, Lieb2002}. However, it is also known that nonlinear interactions cause the wavefunction describing quantum systems to spread around the mean value, creating deviations from MFT on some timescale. The results of quantum effects on observable quantities in the evolution of SFDM is a question of current interest \cite{Sikivie2017, Hertzberg2016, KoppFragkosPikovski2021, Chakrabarty:2017fkd, chakrabarty2021, Lisanti, Lentz2019, Lentz2020}. 
For interacting theories, genuine quantum effects such as squeezing can arise even on time scales before the MFT breaks down \cite{TanasKielich1983, ParkinsWalls1998, JohnssonHaine2007,  KoppFragkosPikovski2021}. 

The classical approximation can be achieved by replacing the field operators with their expectation values, hence ``mean" field theory \cite{Ball}. If the root variance of the field operators is small compared to their mean values then the quantum correction to MFT will be small, and can generally by safely ignored \cite{Eberhardt2021}. In this work, we simulate a toy quantum system, using a highly parallelizable algorithm, and compare them to their MFT approximations. The code used to simulate these systems is publicly available at \href{https://github.com/andillio/QIBS}{https://github.com/andillio/QIBS}.

There is no unique way to measure the deviation of the quantum solution from the classical solution. Therefore, we will parameterize the divergence from a classical description in a number of ways. We will measure the $Q = \sum_i \braket{\delta \hat a_i^\dagger \delta \hat a_i}/n_{tot}$ parameter which is a proxy for the leading order corrective term to the classical field equations \cite{Eberhardt2021}. We will also measure the principle eigenvalue of the $M_{ij} = \braket{\hat a_i^\dagger \hat a_j}$ matrix, this allows us to evaluate how well the Penrose-Onsanger (PO) criterion is satisfied \cite{Penrose1956}. Likewise, we will track the degree to which the quantum state has been squeezed following \cite{KoppFragkosPikovski2021}. We track these quantities as we increase the total number of particles, defining some threshold values in order to define a ``quantum breaktime" at which point the quantum corrections to the MFT are no longer small. 

We will investigate the behavior of different initial quantum states as a function of occupation number. These initial states will include: number eigenstates, which are known to deviate from a single classical field even at large occupation \cite{Sikivie2017, Hertzberg2016, chakrabarty2021}; coherent states, which are the ``most classical states", being initially exactly described by MFT; and field number states, which initially satisfy the PO criterion but have zero initial field operator expectation. Number eigenstate initial conditions are of interest given the existing body of work which has studied them in the context of scalar field dark matter \cite{Sikivie2017, Hertzberg2016, chakrabarty2021}.
Likewise, coherent state initial conditions are of particular interest as it is expected that scalar field dark matter created via the misalignment mechanism will be described by a coherent state at early times \cite{ABBOTT1983133, Preskill:1982}. 

A number of calculations of this convergence and its implications for the quantum breaktime of scalar field dark matter exist in the literature \cite{Sikivie2017, Chakrabarty:2017fkd, Hertzberg2016}.
Previous investigations of the behavior of number eigenstates initial conditions with nonlinear Hamiltonians have concluded that MFT theory admits quantum corrections on a timescale that is set by the classical dynamical time of the interacting system and that increasing occupation number does not result in a convergence to the classical solution \cite{Sikivie2017}. 
We also find that this is true for number eigenstates. However, coherent state and field number state initial conditions converge to the classical solution as the total occupation number is increased. We find that this convergence is approximately consistent with the logarithmic enhancement in occupation number predicted in \cite{Hertzberg2016, Albrecht2014} in systems most similar to scalar field dark matter, i.e. those with quadratic dispersion relations and long range non linear interactions. For systems with linear dispersions and no long range interactions, more similar to the ones studied in \cite{Sikivie2017, Erken2012, Hertzberg2016}, we find that the convergence to the MFT is faster than a logarithmic enhancement. These results may initially appear in contradiction to those in \cite{Sikivie2017}, where it was claimed that even coherent state initial conditions do not match the MFT evolution after one dynamical time scale. However, while this is true for the relatively small initial occupations numbers chosen, the convergence of the solution to the MFT as occupations are increased cannot be ruled out by looking at the evolution of a single set of initial conditions. 

The paper is organized as follows. In Section \ref{sec:Background}, we discuss the relevant physics background. Sections \ref{sec:numerics}, contains a description of our numerical method. We then show our results in Section \ref{sec:results}. Finally, Section \ref{sec:conclusion} contains discussion of these results.

\section{Background} \label{sec:Background}

\subsection{Interacting scalar systems}

The dynamics of the system are described by its Hamiltonian. We will use the following

\begin{equation} \label{Ham}
    \hat H = \sum_j^M \omega_j \hat a_j^\dagger \hat a_j + \sum_{ijkl}^M \frac{\Lambda_{kl}^{ij}}{2} \hat a_k^\dagger \hat a_l^\dagger \hat a_i \hat a_j \, .
\end{equation}

This describe interactions of a non-relativistic scalar field on $M$ modes. $\hat a_j$ is the annihilation operator of mode $j$, it is also the mode space field operator. $\omega_j$ is the kinetic energy associated with mode $j$. A wide variety of systems can be represented by appropriately choosing the weights in the interaction constant $\Lambda_{kl}^{ij}$. 

We can construct a spatial field operator as follows

\begin{align} \label{psi2a}
    \hat \psi (x) = \sum_j \hat a_j u_j(x) \, .
\end{align}

Where $u_j(x)$ is the $j$th mode function represented in the position basis. Throughout this work we will use plane wave mode functions, meaning that $\hat \psi(x)$ and $\hat a_j$ will be related by Fourier transform. This means that $\hat a_j$ can be thought of as the momentum space field operator. 

We will define the interaction terms in terms of a long range interaction constant $C$ and contact interaction constant $\Lambda_0$ giving

\begin{equation} \label{Lambda}
    \Lambda^{ij}_{pl} = \left( \frac{C}{2(p_p - p_i)^2} + \frac{C}{2(p_p - p_j)^2} + \Lambda_0 \right) \delta^{ij}_{pl} \, .
\end{equation}

Where negative constants define attractive potentials and positive constant repulsive ones. $p_j$ is the momentum of the $j$th mode. $\delta^{ij}_{pl}$ is the Kroneker delta, meaning our Hamiltonian explicitly conserves particle number and momentum. 

The evolution of an arbitrary quantum state $\ket{\phi (t)}$ is given by the Schr\"odinger equation

\begin{align}
    \partial_t \ket{\phi(t)} = -i \,  \hat H \ket{\phi(t)} \, .
\end{align}

We have set $\hbar \equiv 1$. 

The evolution of our field operator $\hat a$ can be found using the Heisenberg equation of motion\cite{Ball}, giving the following equations of motion

\begin{align} \label{eqnMotion}
    \partial_t \hat a_p &= i [\hat H, \hat a_p]  
    = -i\left[ \omega_p \hat a_p + \sum_{ijl} \Lambda^{ij}_{pl} \hat a_l^\dagger \hat a_i \hat a_j \right] \, .
\end{align}

The mean field approximation then simply replaces the operators in this equation with their mean values. i.e. $\hat a_p \rightarrow \braket{\hat a_p} \equiv a_p$, giving the classical equations of motion

\begin{align} \label{clEqnMotion}
    \partial_t  a_p &=  -i\left[ \omega_p  a_p + \sum_{ijl} \Lambda^{ij}_{pl} a_l^\dagger a_i a_j \right] \, .
\end{align}
Note that we could have also performed this procedure in the position basis. In general, the classical theory will assume the following is true at all times

\begin{align} \label{clAssumption}
    \braket{\hat a_i^\dagger \hat a_i} = |a_i|^2 \, ,
\end{align}

meaning that the occupation densities can be reconstructed using the classical field. Therefore, we will in general choose our initial conditions such that the classical MFT approximation of a system has $|a_i| = \sqrt{n_i}$. Where $n_i = \braket{\hat a_i^\dagger \hat a_i}$ is the initial expectation value of the $i$th mode occupation.

\subsection{Quantum states}

In this work we find it easiest to represent quantum states in the number eigenstate basis. Physically, these states represent a fixed mode occupation number. In general, a number eigenstate is described by its set of mode occupations $\set{n}$ and is defined as

\begin{align} \label{numEigenstate}
    \ket{\set{n}} &= \ket{n_1, n_2, \dots, n_{M}} \, , \\
    \hat a_j^\dagger \hat a_j \,  \ket{\set{n}} &= n_j \, \ket{\set{n}} \, .
\end{align}

Initial conditions consisting of a single number eigenstate will be among the initial conditions we simulate in this paper. It should be pointed out that it has been demonstrated that a number eigenstates does not converge to a single field classical description even in the high occupation number limit \cite{Sikivie2017, Hertzberg2016}. 

We will also be interested in coherent states. Physically, these states represent a system where the vacuum state is displaced, the $i$th mode occupation number is Poisson distributed with expectation value $|z_i|^2$. The coherent state is described by a vector of complex numbers $\Vec{z} \in \mathbb{C}^M$. In the number eigenstate basis, the coherent state is written 

\begin{equation} \label{coherentStates}
    \ket{\Vec{z}}_C = \bigotimes_{i=1}^M \exp \left[ -\frac{|z_i|^2}{2} \right] \sum_{n_i=0}^\infty \frac{ z_i^{n_i}}{\sqrt{n_i!}} \ket{n_i} \, .
\end{equation}

The coherent state is an eigenvector of the $\hat a_i$ operator with eigenvalue $z_i$, i.e. $\hat a_i \ket{\Vec{z}} = z_i \ket{\Vec{z}}$. This state is generally considered the ``most classical state". Note that it satisfies the classical assumption in equation \ref{clAssumption}.

Finally, we will be interested in the field number states. Physically, these represent systems with fixed total occupation number but where the mode occupations are multinomial distributed across the modes with the probability of the $i$th mode given $|z_i|^2/n_{tot}$, where $n_{tot} = \sum_i n_i$ is a sum over the mode expectations. Like the coherent state, the field number state is described by a vector of complex numbers $\Vec{z} \in \mathbb{C}^M$. In the number eigenstate basis the field number state is written 

\begin{align} \label{fieldNumState}
    \ket{\Vec{z}}_f = \sum_{\set{n}} \sqrt{n_{tot}!} \bigotimes_{i=1}^M \frac{ z_i^{n_i}}{\sqrt{n_i!}} \ket{n_i} \, .
\end{align}

Note that the field number state also satisfied the classical assumption in equation \ref{clAssumption}. 

\begin{widetext}

\begin{figure*}
	\includegraphics[width = .97\textwidth]{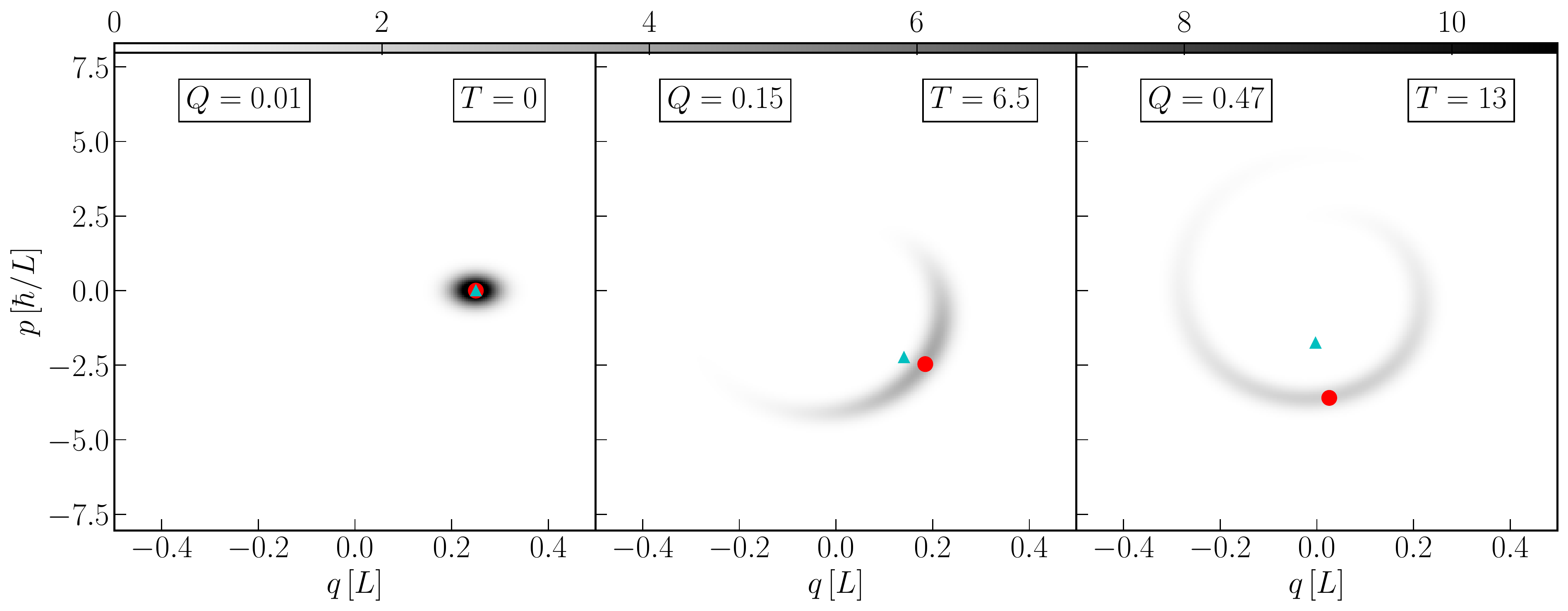}
	\caption{ Here we plot the Husimi distribution of a single mode evolving in a quartic nonlinearity. The mean field value predicted by the classical field theory is shown as a red dot, the true mean field value is shown as a cyan triangle. The time of each snapshot is shown in the top right of each subplot. Overtime the wavefunction spreads due to the nonlinearity resulting in a discrepancy between the classical and actual value of the mean field. This wavefunction spreading is parameterized by $Q$ which is shown in the top left of each subplot. The middle plot show the time defined as the quantum breaktime, at this point the wavefunction has already undergone significant phase diffusion. }
	\label{fig:qualConv}
\end{figure*}

\subsection{Correction terms and Q parameter}

Equation \eqref{clEqnMotion} was achieved by taking the expectation value of equation \eqref{eqnMotion} and then making the approximation that $\partial_t \braket{\hat a_p} = \braket{f(\hat a_p)} \approx f( \braket{\hat a_p} )$. This approximation can be restored to an equality by adding higher moment correction terms as in \cite{Eberhardt2021}

\begin{align} \label{series}
    \partial_t \braket{\hat a_p} &= \braket{f(\hat a_p)} = f( \braket{\hat a_p} ) + \sum_{ij} \braket{\delta \hat a_i^\dagger \delta \hat a_j} \frac{\partial^2}{\partial \braket{\hat a_i^\dagger} \partial \braket{\hat a_j}} f( \braket{\hat a_p} ) + \dots  \, .
\end{align}
\end{widetext}

Where $\delta \hat a_i = \hat a_i - \braket{\hat a_i}$ The leading order correction term is proportional to the second moments of the field operators and the second derivative of the time evolution function with respect to the field expectation. Therefore, we can approximate the average size of the correction compared to the leading order term by defining the $Q$ parameter

\begin{align} \label{Qparam}
    Q \equiv \sum_j \frac{\braket{\delta \hat a_j^\dagger \delta \hat a_j}}{n_{tot}} \, .
\end{align}
This measures how well the quantum distribution is localized around the classical field value. Over time, nonlinear evolution will spread the wavefunction causing quantum correction terms to become relevant to the evolution of the mean field value. When this parameter is no longer small we do not expect the classical field equations to accurately describe the quantum evolution. 

The convergence properties of this parameter, when compared to other classicality criteria, is interesting for a number of reasons. This parameter works well when the classical field theory can be reasonably phrased in terms of the expansion in equation \eqref{series}, where each subsequent term is small compared to the preceding ones, as is the case for coherent state initial conditions with large occupation numbers. Likewise, the computational complexity of calculating $Q$ scales only linearly with mode number $M$. Finally, this parameter can be calculated in computationally inexpensive extensions of the MFT such as \cite{Eberhardt2021}. 

Cosmic scalar field dark matter represents a system of interacting Bosons expected to start in coherent state initial conditions, and have a large number of relevant modes. The $Q$ parameter is useful in these circumstances and therefore understanding its properties using the toy models here is of particular interest. 

\subsection{Penrose-Onsager criterion}

The Penrose-Onsager (PO) criterion is satisfied when we can write the following \cite{Penrose1956}

\begin{equation} \label{POcriterion}
    \braket{\hat a^\dagger_i \hat a_j} = \vec{z}_i^\dagger \vec z_j \, .
\end{equation}

That is that the second moment matrix $M_{ij} \equiv \braket{\hat a^\dagger_i \hat a_j}$ can be written as an outer product of a single field $\vec z$. 

We will test how well this conditions is satisfied by looking at the eigenvalues of $M_{ij}$. When the PO criterion is satisfied there will be a single nonzero ``principal" eigenvalue, $\lambda^p$ equal to the squared norm of $\vec z$. Where $\vec z^* / \sqrt{\sum_i|z_i|^2}$ is the corresponding principal eigenvector, $\vec \xi^p$.  When the system is well described by the classical theory we expect that the principal eigenvalue is very close to $n_{tot}$ \cite{Leggett2001}. If the principal eigenvalue deviates too far from $n_{tot}$ a single classical field description is insufficient.

It should be noted that satisfying the PO criterion does not imply that the conjugate of the principal eigenvector obeys the classical field equations of motion though this is often, at least approximately, the case, see for example figure \ref{fig:POparam}.

Often when simulating a physical system the mode occupation or spatial densities are of specific interest. The PO criterion is a useful measure of classicality because when it is satisfies this implies that there exist a single field which captures these occupation numbers. Computationally, this requires solving matrix eigenvalues and therefore has cubic scaling with the mode number, $M$.

\subsection{Squeezing} \label{sq_subsection}
Squeezing of a quantum state is present if the uncertainty of an operator becomes smaller than for a reference vacuum state. Squeezing is often considered a signature of non-classicality, or ``quantumness'', especially in the context of quantum optics \cite{TanasKielich1983, drummond2013quantum}.
Let us consider some operator $\hat O$ with $[\hat O, \hat O^\dagger] = 1$. Then a Hermitian operator, a so-called quadrature, can be defined via
\begin{equation}
     \hat X_\theta = \hat O e^{-i \theta}+\hat O^\dagger e^{i \theta} \,, 
\end{equation}
where $\theta$ is a parameter.
This has variance
\begin{align} 
    \mathrm{Var}(\hat  X_\theta) & =  1+ 2 \langle \delta \hat O^\dagger \delta \hat O\rangle   \notag\\
    & \qquad+ e^{-2i \theta} \mathrm{Var}(\hat O) + e^{2i \theta} \mathrm{Var}(\hat O^\dagger)
\end{align}
where $\mathrm{Var}(\hat  O) \equiv \langle \hat O^2 \rangle - \langle \hat O \rangle^2 $. 
%
%
At the angle $\theta_-$ the variance $V_- \equiv \mathrm{Var}(\hat  X_{\theta_-})$ is minimised and given by
\begin{equation}\label{generalVminus}
    V_-(t) = 1+ 2 \langle \delta \hat O^\dagger \delta \hat O\rangle  - 2 |\mathrm{Var}(\hat O)|\,.
\end{equation}
If the quantum state is the vacuum, or a coherent state, then $V_- =1$. If $V_- < 1$ the state is said to be squeezed \cite{drummond2013quantum}. 
In general a quantum state can be squeezed with respect to multiple operators. 
Here will focus on the squeezing of individual mode operators $\hat a_i$, and the annihilation operator that destroys particle in the state corresponding to the principal component of $\hat M$.
\begin{align} \label{a_op}
 \hat a^{p}  \equiv   \sum_k  \xi^p_{k} \,  \hat a_k\,.
\end{align}
The significance of $\hat a^{p}$ is the following. 
If the PO criterion is valid, then a single field description should be possible, and squeezing should be encapsulated by $\hat a^{p}$.
In the case where $\lambda^p=n_{tot}$, and for an initial coherent state, the squeezing of SFDM can be studied analytically. Formulas for $V_-(t)$ and $\theta_-(t)$, the time of onset of squeezing, as well as the time and magnitude of maximal squeezing were obtained in \cite{KoppFragkosPikovski2021}.
We can define the onset of squeezing $t_{\rm sqz}$ to be when $V_-(t_{\rm sqz}) = 0.8$ which is reached at $t_{\rm sqz} \simeq 0.05 (n_{\rm tot} |\chi|)^{-1}$, where $n_{\rm tot} \chi =  n^{-1}_{\rm tot}\sum_{ijkl} \tfrac{1}{2}\Lambda^{ij}_{kl} z^*_k z^*_l z_i z_j$  is the mean interaction energy per occupation. We thus have
\begin{equation}
    t_{\rm sqz} = 0.05 \frac{ n_{\rm tot}}{|\sum_{ijkl} \tfrac{1}{2}\Lambda^{ij}_{kl} z^*_k z^*_l z_i z_j|} \,.
\end{equation}
Maximal squeezing $V_-(t_{\rm max}) = V_{\rm min} \simeq n_{\rm tot}^{-1/3}$ is obtained at $t_{\rm max} \simeq 0.5 t_{\rm sqz} n^{1/6}_{\rm tot}$.
The time evolution of $V_{-}(t)$ can for $n_{\rm tot} >20$ be accurately approximated by 
\begin{align} 
    V_-(t) &\simeq 1 - 4 \tau s + 8 \tau^2  + \frac{8 \tau^3(5+12 \tau^2)}{n_{tot} s}  - \frac{16 \tau^4}{n_{tot}}\nonumber\\
    s & \equiv \sqrt{1+ 4 \tau ^2} \quad, \qquad\tau \equiv   n_{\rm tot}  | \chi | t  \,. \label{VminusBajer}
\end{align}
At early times we thus get
\begin{align} 
    V_-(t) &= 1 - 4 \tau + \mathcal{O}(t^2) \,.
\end{align}

\section{Numerical Implementation} \label{sec:numerics}

\subsection{Classical field}

The classical field equations of motion are obtained by replacing the operators in equation \eqref{Ham} with their expectation values. The classial field, $a^{cl}_p$ then evolves as  

\begin{align} \label{classicalFieldEqns}
    \partial_t a^{cl}_p = -i\left[ \omega_p a^{cl}_p + \sum_{ijl} \Lambda^{ij}_{pl} a^{cl\dagger}_l a^{cl}_i a^{cl}_j \right] \, .
\end{align}

We integrate this equation using the CHiMES package available at \href{https://github.com/andillio/CHiMES}{https://github.com/andillio/CHiMES} described in \cite{Eberhardt2021}.

The initial conditions used to approximate each quantum state are given as 
\begin{enumerate}
    \item Number eigenstate: $a^{cl}_p \biggr\rvert_{t=0} = \sqrt{ n_p }$. 
    \item Coherent state: $a^{cl}_p \biggr\rvert_{t=0} = z_p$. 
    \item Field number state: $a^{cl}_p \biggr\rvert_{t=0} = z_p$.
\end{enumerate}

\subsection{Quantum field}

The full code repository for the simulation and data analyses of the quantum simulations performed here is publicly available at \href{https://github.com/andillio/QIBS}{https://github.com/andillio/QIBS}.

The evolution of the quantum system is solved by integrating Schr\"odinger's equation 

\begin{equation} \label{Schr}
    \partial_t \ket{\phi} = -i \hat H \, \ket{\phi} \, .
\end{equation}

Where the Hamiltonian is given in equation \eqref{Ham}. 

For most problems with $M > 1$ the dimensionality of the total Hilbert space of the system, $\mathcal{H}_T$, which contains all the number eigenstates relevant to the evolution of our state, is quite large. Let $D_T = \mathrm{Dim}[\mathcal{H}_T]$. Directly integrating equation \eqref{Schr} would be $\mathcal{O}(D_T^2)$. However, since our Hamiltonian conserves particle number and momentum, we can efficiently simulate the system by partitioning the total Hilbert space into sub-spaces that can be solved in parallel.

We can write any state as a sum over number eigenstates as

\begin{equation} \label{sum_of_nums}
    \ket{\phi} = \sum_{ j } c_j \ket{\set{n}_j} \, .
\end{equation}

We truncate this sum such that $\braket{\phi | \phi} \ge 0.999$.

We can then rewrite equation \eqref{Schr} as

\begin{equation} \label{schr_n}
    \partial_t \ket{\phi} = \sum_j \partial_t \ket{\set{n}_j}  = -i \sum_{ j } c_j \, \hat H \ket{\set{n}_j} 
\end{equation}

Because of the number and momentum conserving properties of our Hamiltonian, the time evolution of any given number eigenstate, $\ket{\set{n}}$ can only ever evolve into a superposition of number eigenstates with the same total number of particles and total momentum. This means the dimensionality of the Hilbert space relevant to the evolution of any given term in equation \eqref{schr_n} is much smaller than the dimensionality of the total Hilbert space. 

For any number eigenstate $\ket{\set{n}}$ then, with occupations $\set{n}$, we can define such a ``special Hilbert space", $\mathcal{H}^{\set{n}} \subseteq \mathcal{H}_T$, which contains all the terms $\ket{\set{n'}} \in \mathcal{H}_T$ such that 

\begin{align}
    \sum_{n'_k \in \set{n'}} n'_k &= \sum_{n_k \in \set{n}} n_k \, \textrm{, and} \\
    \sum_{n'_k \in \set{n'}} k \, n'_k &= \sum_{n_k \in \set{n}} k \, n_k \, .
\end{align}

That is, when our mode functions correspond to the momentum eigenstates, a given special Hilbert space contains all the number eigenstates with the same total number of particles and net momentum. We can therefore uniquely identify a subspace by its net momentum and total particle number. Every number eigenstate has a definite net momentum and total particle number, and therefore every state is in exactly one special Hilbert space. Thus, the special Hilbert spaces partition the total Hilbert space. Letting $D^j = \mathrm{Dim}[\mathcal{H}^j]$, we can say that $D_T = \sum_{\textrm{$\mathcal{H}^j \in \mathcal{H}_T$}} D^j$.

For each term, $\ket{\set{n}_j}$, in the sum on the right side of equation \eqref{schr_n} we can associate a special Hilbert space $\mathcal{H}^j$. We can then represent the Hamiltonian, $\hat H^j$ and number eigenstate term in this smaller special Hilbert space.

\begin{equation}
    \hat H_{lm}^j = \braket{\set{n}_l | \hat H | \set{n}_m}
\end{equation}

for all $\ket{\set{n}_l}, \ket{\set{n}_m} \in \mathcal{H}^j $. This corresponds to the Hamitlonian projected into the $j$th special Hilbert space. We can also write a projection operator for each special Hilbert space as

\begin{equation}
    \hat P^j = \sum_{\ket{\set{n}} \in \mathcal{H}^j} \ket{\set{n}} \bra{\set{n}} \, .
\end{equation}

We can rewrite our state in terms of projections into special Hilbert spaces and time dependant complex weights $\set{w}$ as

\begin{align}
    \ket{\phi(t)} &= \sum_j \hat P^j \ket{\phi(t)} \, , \\
    &= \sum_j \ket{\phi^j(t)} \, , \\
    &= \sum_j \sum_{\ket{\set{n}_k} \in \mathcal{H}^j} w^j_k(t) \ket{\set{n}_k} \, .
\end{align}

The initial values of $w^j_k$ are given 

\begin{equation}
    w^j_k(t=0) = \braket{\set{n}_k | \phi^j \, (t=0)}
\end{equation}

We can then rewrite equation \eqref{schr_n} as

\begin{equation}
    \partial_t \ket{\phi} = -i \sum_j \hat H^j \hat P^j \ket{\phi} \, .
\end{equation}

Where the evolution each term in the sum is completely independent of the evolution of the other terms. We can therefore solve for each term in parallel.

\begin{widetext}

We integrate the weights $\vec w^j_i$ using a second order integrator of the Schr\"odinger equation, i.e. we take weights at a time $t$ to a time $t + \Delta t$ via

\begin{equation} \label{Q_int}
    w^j_k(t + \Delta t) = w_k^j(t) - i \Delta t \sum_m \hat H_{km}^j w^j_m(t) - \Delta t^2 \sum_l \hat H_{kl}^j \sum_m \hat H_{lm}^j w^j_m(t) \, .
\end{equation}

\end{widetext}

Using the fact that the special Hilbert spaces partition the total Hilbert space we can say the computational complexity of integrating equation \eqref{Q_int} is $\mathcal{O}(\sum_j (D^{j})^2) \le \mathcal{O}(D_T^2)$. 

However, it should be noted that the number of nonzero values in a given row or column of the Hamiltonian projected into any special Hilbert space, $\hat H^j_{kl}$, can be no greater than the number of interaction terms in equation \eqref{Ham}, i.e. $M^4$. This means that for large enough systems, i.e. when $M^4 \lesssim D^j$ the matrix is sparse and the scaling becomes $\mathcal{O}(\sum_j M^4 D^{j})$.

Finally, the total state $\ket{\phi\,(t)}$ can be recovered by summing over the special Hilbert spaces and associated weights as follows

\begin{equation}
    \ket{\phi (t)} = \sum_j \sum_k w^j_{k}(t) \ket{\set{n}_k} \, .
\end{equation}

\section{Results} \label{sec:results}
\subsection{Test Problems}

In order to test how increasing occupation numbers effect the duration of an accurate classical field approximation for a variety of systems, we run the following simulation parameters and initial conditions. The initial conditions used are

\begin{enumerate}
    \item \textbf{Number eigenstates.} We simulate the number eigenstate  $\ket{0,\, 2r, \, 2r, \, 1r,0}$. 
    \item \textbf{Coherent state.} We simulate the coherent state $\vec z = (0,\,\sqrt{2r}\, e^{i\theta_1},\,\sqrt{2r}\,e^{i\theta_2} ,\,\sqrt{1r}\,e^{i\theta_3},\,0)$. Phases are drawn from a uniform random distribution, $\theta_i \sim U[0, 2\pi)$ with fixed random seed. The sum in equation \ref{coherentStates} is truncated such that $\braket{\vec z| \vec z} \ge .999$. 
    \item \textbf{Field number state.} We simulate the field number state $\vec z = (0,\,\sqrt{2r}\, e^{i\theta_1},\,\sqrt{2r}\,e^{i\theta_2} ,\,\sqrt{1r}\,e^{i\theta_3},\,0)$. Phases are drawn from a uniform random distribution, $\theta_i \sim U[0, 2\pi)$ with fixed random seed. The sum in equation \ref{fieldNumState} is truncated such that $\braket{\vec z| \vec z} \ge .999$.
\end{enumerate}

and simulation parameters

\begin{enumerate}
    \item \textbf{Attractive long range interactions.} Here we set $\omega_j = j^2 / 2$, $\Lambda_0= 0$, and $C = -0.1/r$.
    \item \textbf{Repulsive/Attractive contact interactions.} Here we set $\omega_j = j / r$, $\Lambda_0= \pm 0.1 / r$, and $C = 0$. 
\end{enumerate}
where $r$ here is a scaling parameter equal to the average mode occupation. We choose this scaling parameter so our results can be more easily compared with those in \cite{Sikivie2017}. 

\begin{figure}
	\includegraphics[width = .45\textwidth]{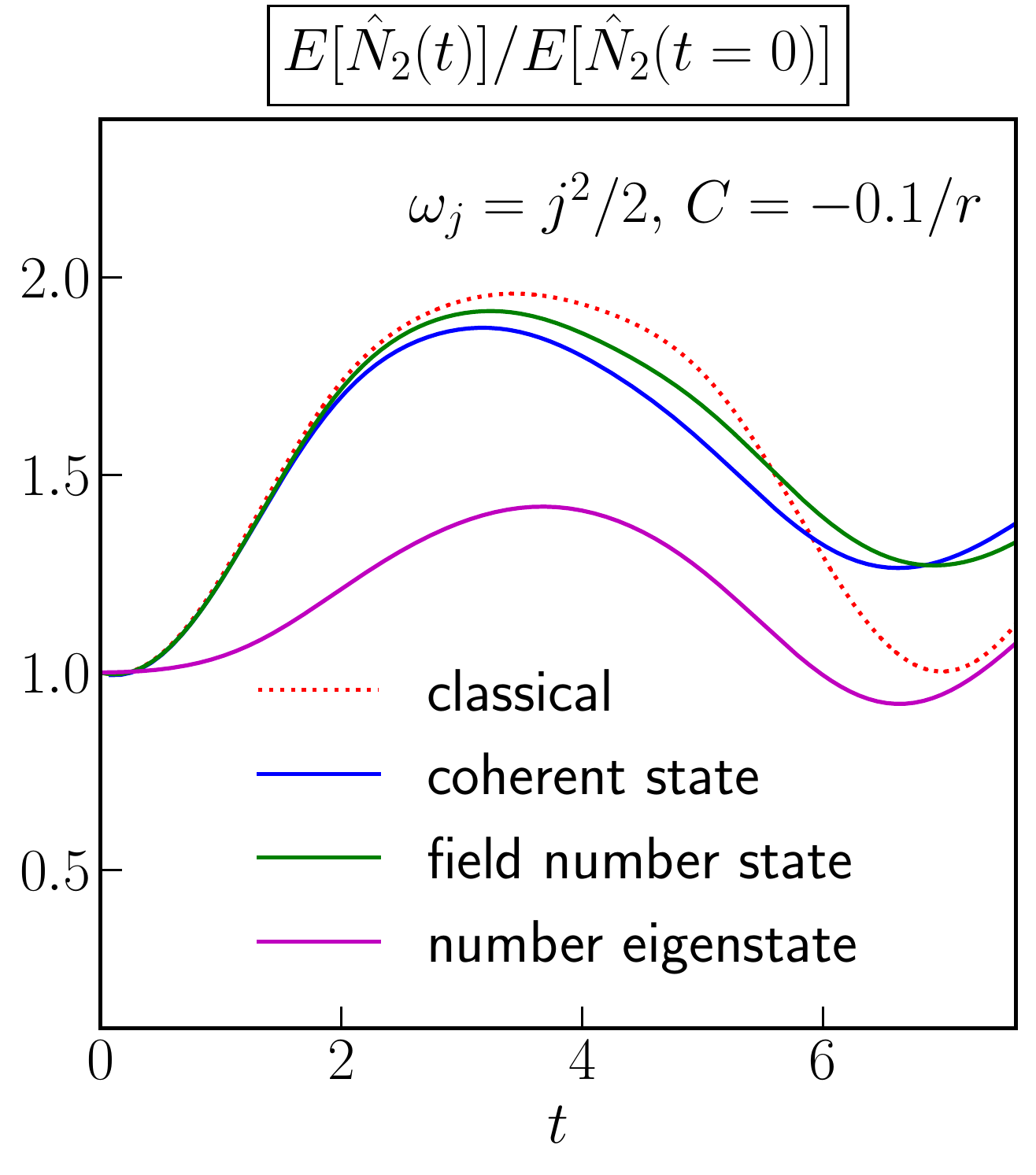}
	\caption{ Here we show the quantum evolution of the occupation number of a particular mode for different initial conditions, shown in solid, compared to the classical evolution, show in dotted red. Notice that the evolution of number eigenstate initial conditions differ from the classical evolution on dynamical timescales, admitting large corrections almost immediately. In contrast, coherent and field number state initial conditions closely track the classical evolution at early times. Here we set $r=11$. }
	\label{fig:ICs}
\end{figure}

In general, we expect that the mode occupation number values of number eigenstates to deviate from the classical evolution on dynamical timescales, as this is simply the timescale on which the system will differ from its own initial conditions. We expect coherent and field number states to adhere to the classical solution for at early times. This can be seen for long range interactions in figure \ref{fig:ICs}.

\subsection{Qualitative convergence}

\begin{figure*}
	\includegraphics[width = .97\textwidth]{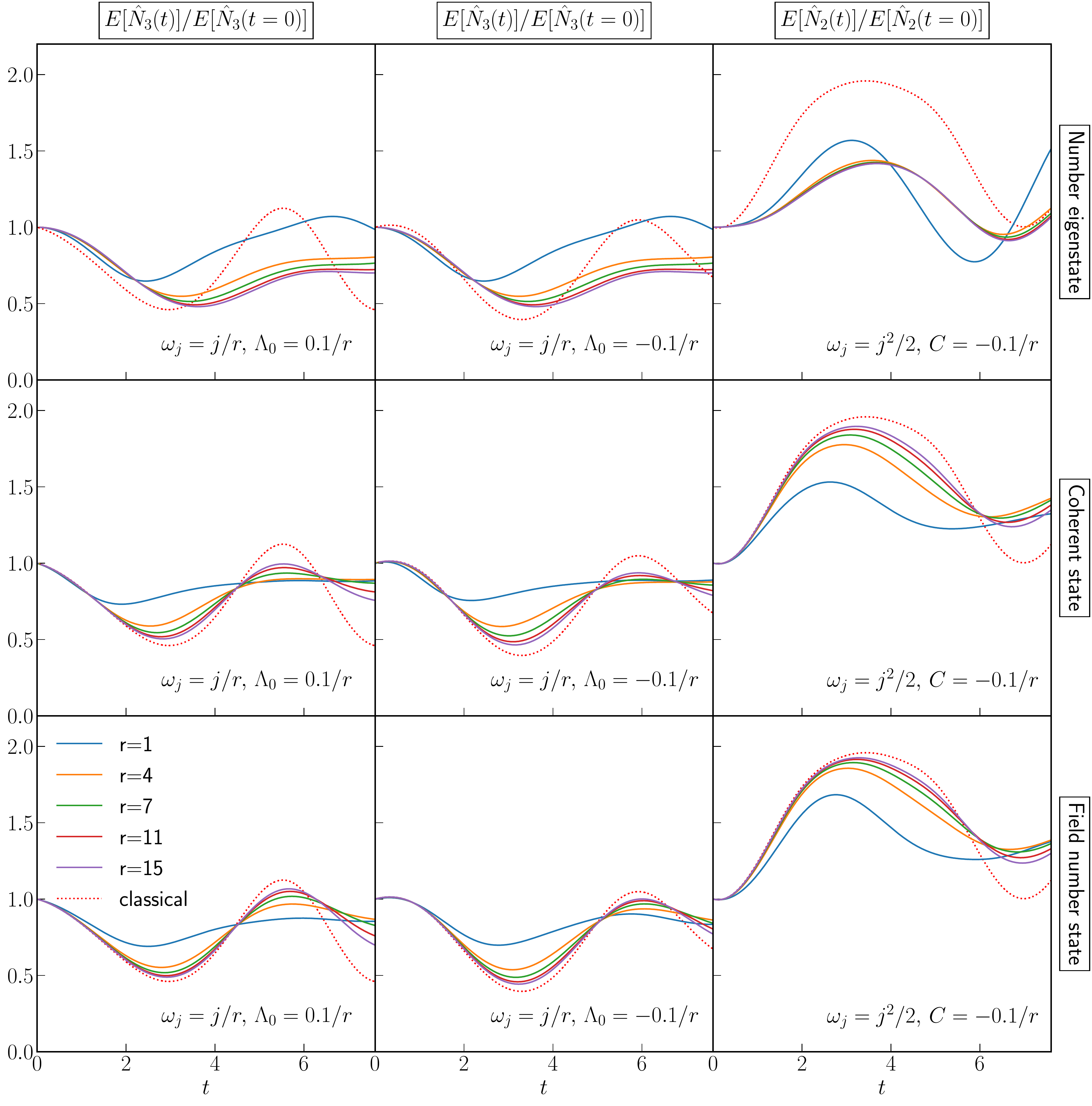}
	\caption{ Here we plot the evolution of the occupation numbers of a given mode for our test problems. Each row represents different quantum state initial conditions, and each column different simulation parameters. We compare the quantum evolution at varying average mode occupation number, $r$, to the classical solution shown in dotted red. We can see that for the coherent states and field number eigenstates that increasing the occupation numbers leads to a quantum evolution which converges to the classical evolution. The number eigenstate does not approach the classical solution regardless of occupation number and differ on the same timescale as it takes for the evolution of the mode occupation number to differ from the initial conditions, the dynamical time. }
	\label{fig:qualConv}
\end{figure*}

We can get a intuitive sense for how increasing the occupation number effects convergence to the classical solution by looking at how the evolution of the mode occupations approach the classical evolution as we change the scaling parameter $r$, compare with Figure 2 in \cite{Sikivie2017}. 

We make this comparison for a given mode for our test problems in Figure \ref{fig:qualConv}. We compare the evolution of the quantum solution occupation numbers and the classical occupation numbers as we change the total particle number occupations, while holding $N(\Lambda_0 + C)$ fixed. We can see that the coherent states and field number states tend to converge towards the classical solution as the occupation numbers are increased. We can see also that the rate of convergence diminishes as occupation number is increased. Notice that increasing the occupation numbers of the number eigenstate does not cause it to approach the classical solution as noted in \cite{Hertzberg2016, Sikivie2017}. A single field description is inadequate for describing number eigenstates, however it was demonstrated in \cite{Hertzberg2016} that an ensemble of fields was an accurate approximation of the quantum evolution.

\subsection{Q Parameter}

\begin{figure}
	\includegraphics[width = .45\textwidth]{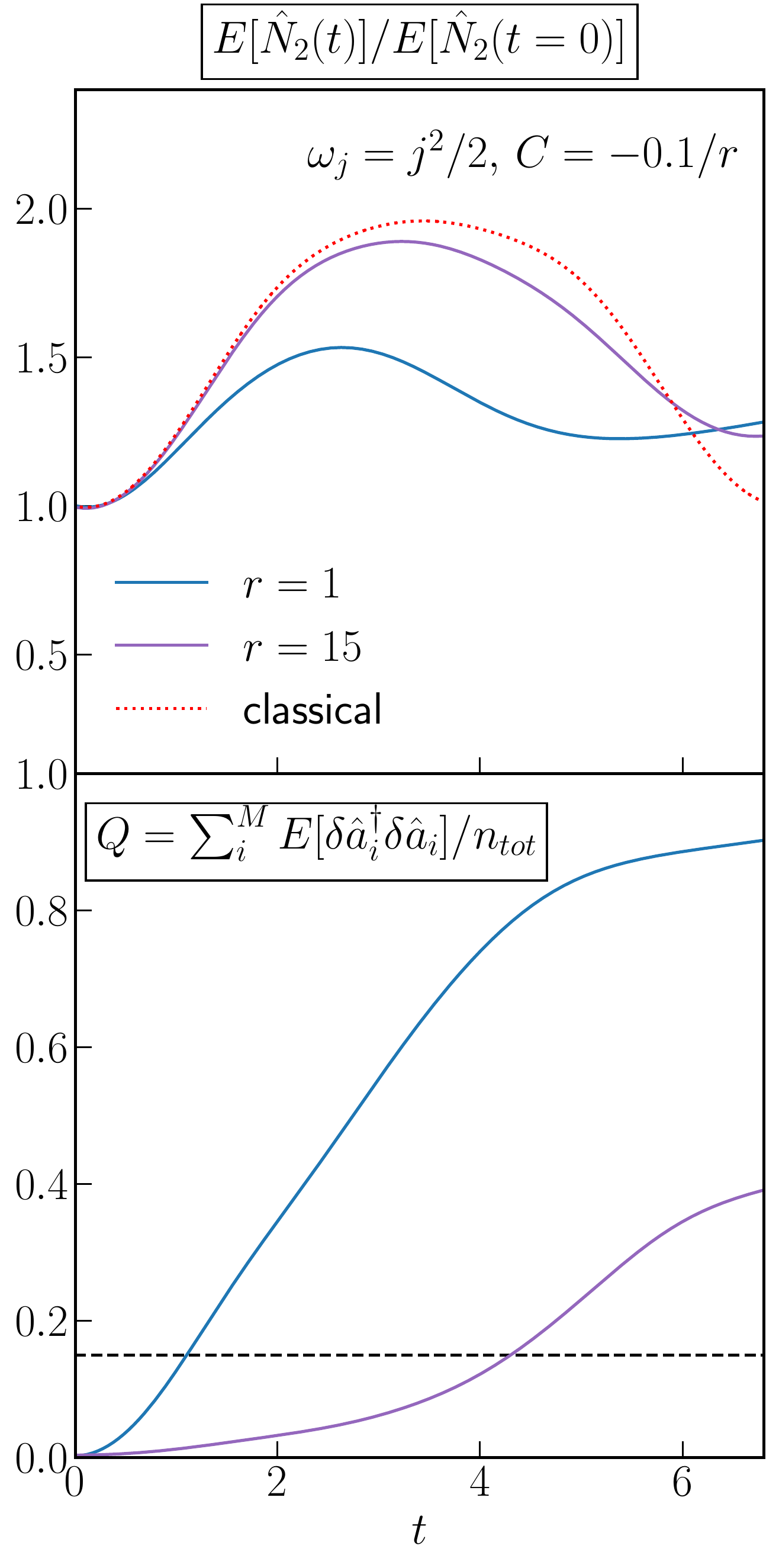}
	\caption{ Here we plot the evolution and effect of the $Q$ parameter for a single mode two systems with long range interactions, quadratic dispersion, and coherent state initial conditions but different average mode occupation number, $r$. The top subplot shows the evolution of one of the occupation numbers of a specific mode. We show the classical evolution in dotted red with the quantum evolutions shown as solid lines. The bottom subplot shows the value of the $Q$ parameter for each system, the dashed black line indicates the breakpoint threshold value as we have defined it. We can see that past the breakpoint for each system the classical occupation number of the plotted mode begins to deviate from the actual quantum results. Consistent with other results we see that larger occupation number systems follow the classical solution longer. }
	\label{fig:Qparam}
\end{figure}

\begin{figure*}
	\includegraphics[width = .97\textwidth]{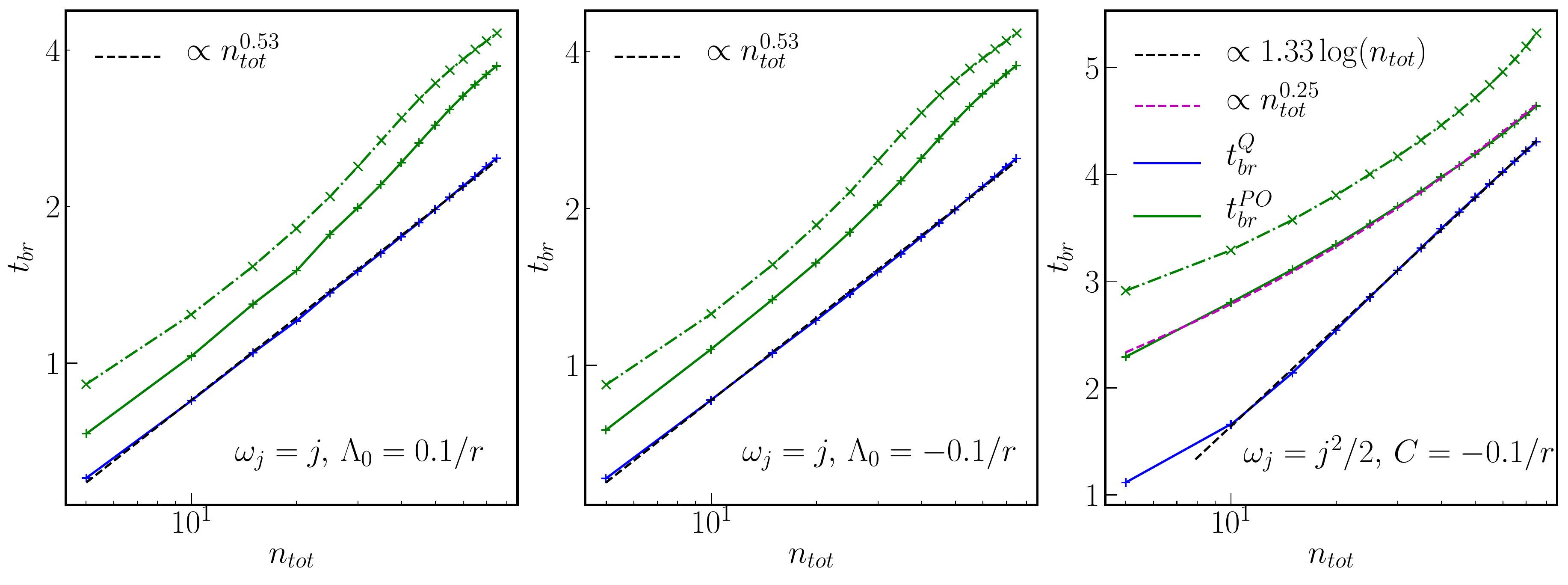}
	\caption{Here we show how the quantum breaktime scales with total particle number. Both coherent and field number states converge to the classical solution for all criteria, as shown by the increasing breaktime as a function of $n_{tot}$. Solid lines and plus markers indicate coherent state initial conditions, dash dotted lines with cross markers indicate field number state initial conditions. Blue lines indicate where $Q(t^Q_{br}) \equiv 0.15$, green lines where $1-\lambda_p(t^{PO}_{br})/n_{tot} = 0.1$. Note that the rightmost plot has a linear vertical axis. The dashed lines are best fits assuming either a logarithmic enhancement or power law scaling of the breaktime with $n_{tot}$. }
	\label{fig:params}
\end{figure*}

The $Q$ parameter gives us a sense of the size of the leading order correction to the MFT relative to the classical terms. This is a measure of how well the quantum distribution of the field value is centered on the classical field value in phase space. By defining a breaktime to occur when these correction terms are no longer sub-leading order we can get a sense of how the quantum evolution converges to the classical. 

Here we simulate coherent state initial conditions which are known to be initially well described by a classical field. The correction terms, and correspondingly the $Q$ parameter are initially $0$. Over time, the interaction term in equation \eqref{Ham} causes the wavefunction to spread from the mean field value leading to an increase in the $Q$ parameter, see figure \ref{fig:Qparam}. We follow the $Q$ parameter until it is no longer $Q \not\ll 1$. We define this occur at $Q(t^Q_{br}) \equiv 0.15$, this choice is somewhat arbitrary and our results do not depend sensitively on reasonable adjustment of this definition.

For all simulation parameters similar to scalar field dark matter (i.e. a quadratic dispersion, long range interactions, and coherent state initial conditions) we observe an approximate logarithmic enhancement in the quantum breaktime as a function of the total particle number, as can be seen by the blue lines in the rightmost plot in figure \ref{fig:params}. Consistent with expectations for chaotic interacting systems in expressed in \cite{Hertzberg2016}. This logarithmic scaling corroborates what we observe in Figure \ref{fig:qualConv}, namely that the rate of converges diminishes as the occupation number is increased.

For all other simulation parameters (i.e those with contact interactions or linear dispersions) we observe that $Q$ grows approximately quadratically. This results in a power law scaling of the breaktime, see figures \ref{fig:params}. 

Note that this parameter is not effective for evaluating the deviation of state where the expectation value of the field operator is always $0$ such as the number eigenstate and field number state. 

\begin{figure}
	\includegraphics[width = .45\textwidth]{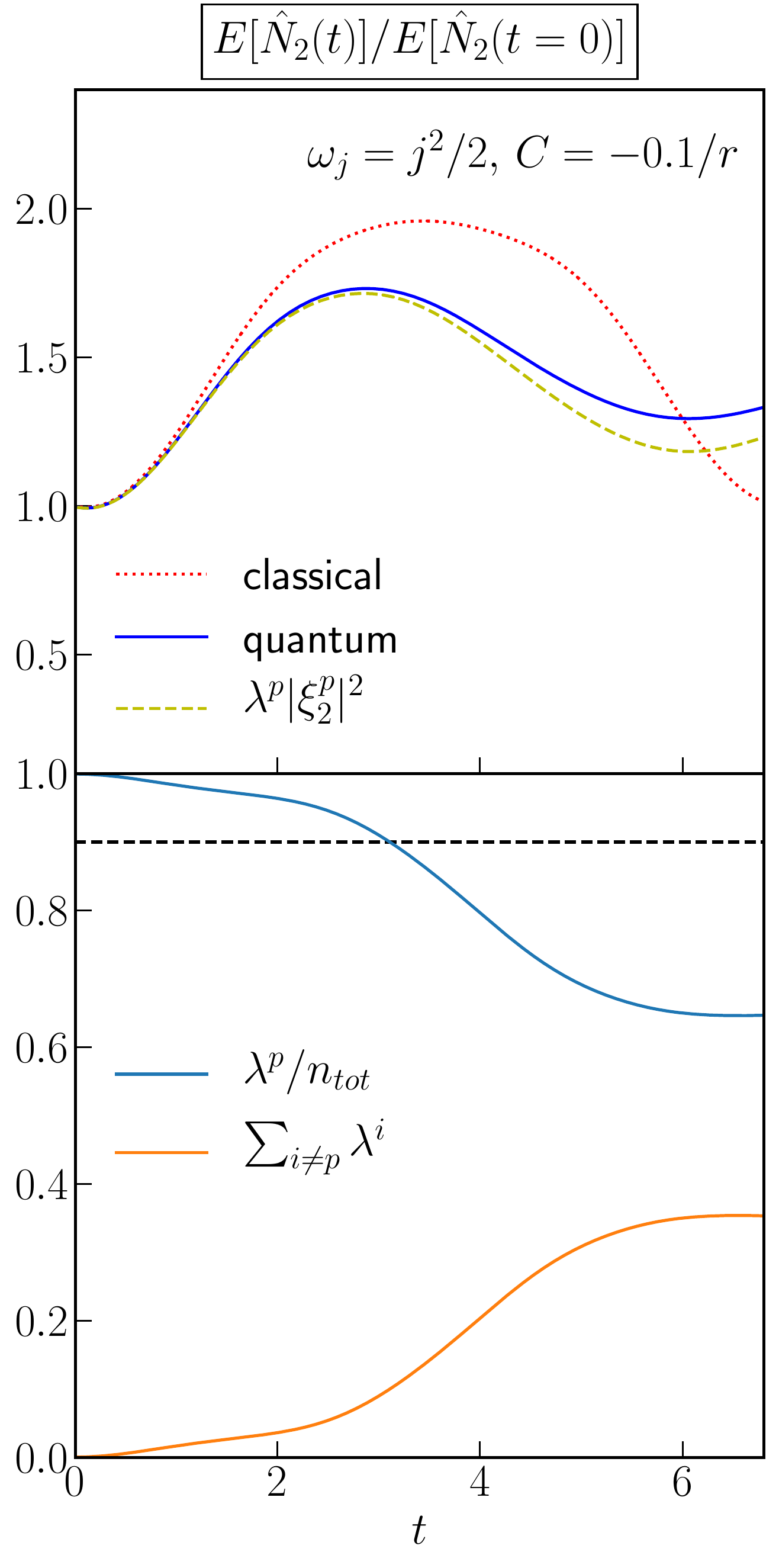}
	\caption{ Here we plot the evolution of the PO criterion for a system with long range interactions, quadratic dispersion, and $r=3$ average mode occupation number. In the top subplot we compare the classical and quantum occupation number of the second mode to the square amplitude of the principle eigenvector multiplied by the principle eigenvalue. The bottom plot shows how the principle eigenvalue, shown in blue, compares to the total number of particles. The dashed black line indicating the breakpoint threshold as we have defined it. We can see that while all three evolutions agree at early times, past the breaktime the classical occupation number deviates both from its true quantum value and from the principle eigenvector. Notice also that even when the PO criterion is approximately satisfied it is not necessarily the case that the principle eigenvector follows the classical field equations of motion. }
	\label{fig:POparam}
\end{figure}

\subsection{PO criterion}

The PO criterion evaluates how well the information in $M_{ij}$ the second order moment matrix is described by a single field. When completely described by a single field, $M_{ij}$ will have a single nonzero eigenvalue. As the quantum evolution deviates from a classical description additional eigenvalues will become nonzero, see for example figure \ref{fig:POparam}. Because the trace of $M_{ij}$ is conserved, we can define a breaktime by following when the size of the principle eigenvalue. 

Here we simulate coherent and field number state initial conditions. Both are known to initially satisfy the PO criterion, i.e. $1 - \lambda_p/n_{tot} = 0$. We follow the evolution of the principle eigenvalue until it is no longer approximately equal to the total number of particles, $\lambda_p \not\approx n_{tot}$. We can use this quantity to define another breaktime and check that it has the same qualitative behavior as our previous defined time. We define this breaktime as $1 - \lambda_p(t^{PO}_{br})/n_{tot} \equiv 0.1$.

The scaling of the breaktime using this definition is consistent with our definition using the $Q$ parameter, see green lines Figure \ref{fig:params}. The fact that there is agreement between these two methods is a good indication that our results are robust to the specific definition of a breaktime and can all be used to define a time at which there exists nonnegligible quantum corrections. A noticeable exception to this is for long range interactions with a quadratic dispersion. In this case, the scaling of the PO criterion is consistent with a power law while the $Q$ parameter appears to scale as $\log(x)$.

Note that the PO criterion is not effective for evaluating the deviation of number eigenstates which start with $1-\lambda_p/n_{tot} \sim 1$. 

\subsection{Squeezing}

\begin{figure}
	\includegraphics[width = .45\textwidth]{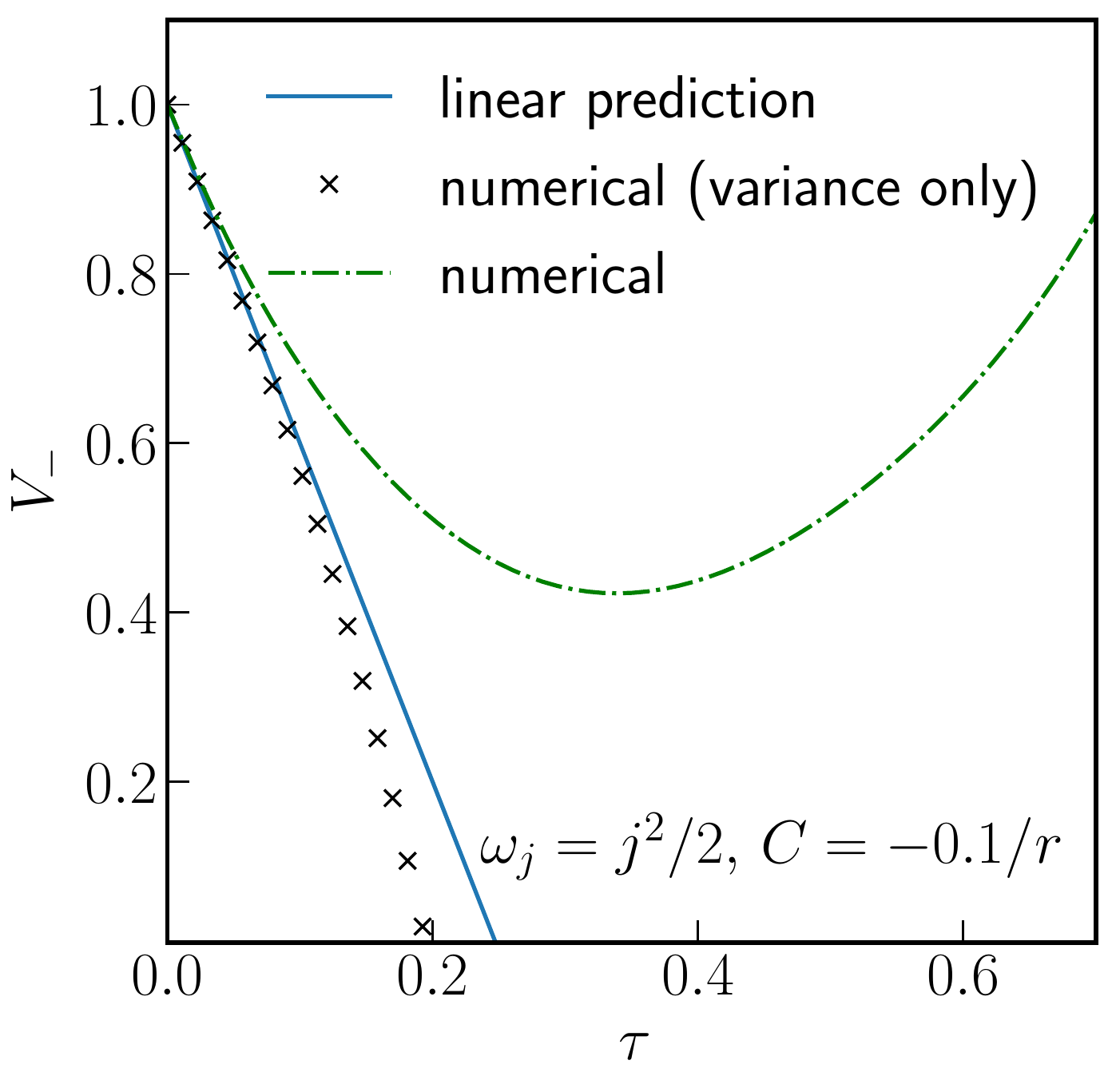}
	\caption{ Here we plot the squeezing of the operator defined in equation \eqref{a_op}, which is the annihilation operator for particles in the field defined by the principle eigenvector of $\hat M$. The solid blue line shows the linear prediction $1 - 4 n_{tot} |\chi| t$, the dashed dotted green line show the simulated value of the squeezing $1 + 2 Cov(\hat a^\dagger, \hat a) - 2|Var(\hat a)|$, and the black x's show the simulated value of the squeezing considering only the variance $1 - 2|Var(\hat a)|$. At very early times, when $\tau \ll 1$, the linear contribution dominates and all three lines agree. The linear prediction eventually fails around $\tau \approx 0.15$. Note that the line corresponding to the contribution of the variance only to squeezing remains close to the linear prediction for longer than the total squeezing. This is because the covariance provides no linear contribution. Here we set $r=11$. }
	\label{fig:sq}
\end{figure}

\begin{figure}
	\includegraphics[width = .45\textwidth]{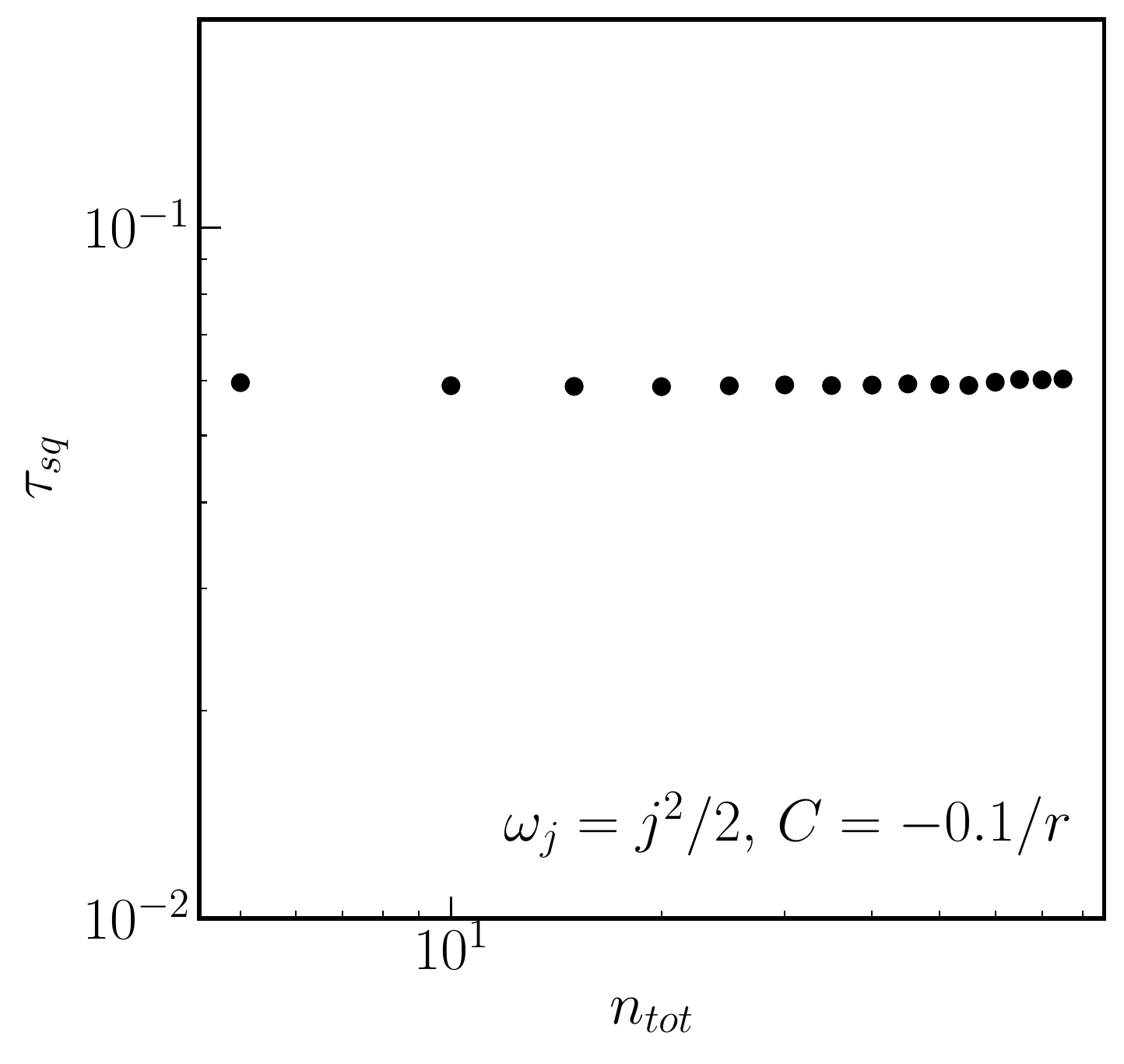}
	\caption{ Here we plot the squeezing time, where $V_-(\tau_{sq}) = 0.8$, as a function of the total particle number. The time is constant, this is a result of the fact that the initial squeezing is proportional to the classical non-linearity which is constant as we increase $n_{tot}$.}
	\label{fig:t_sq}
\end{figure}

\begin{figure}
	\includegraphics[width = .45\textwidth]{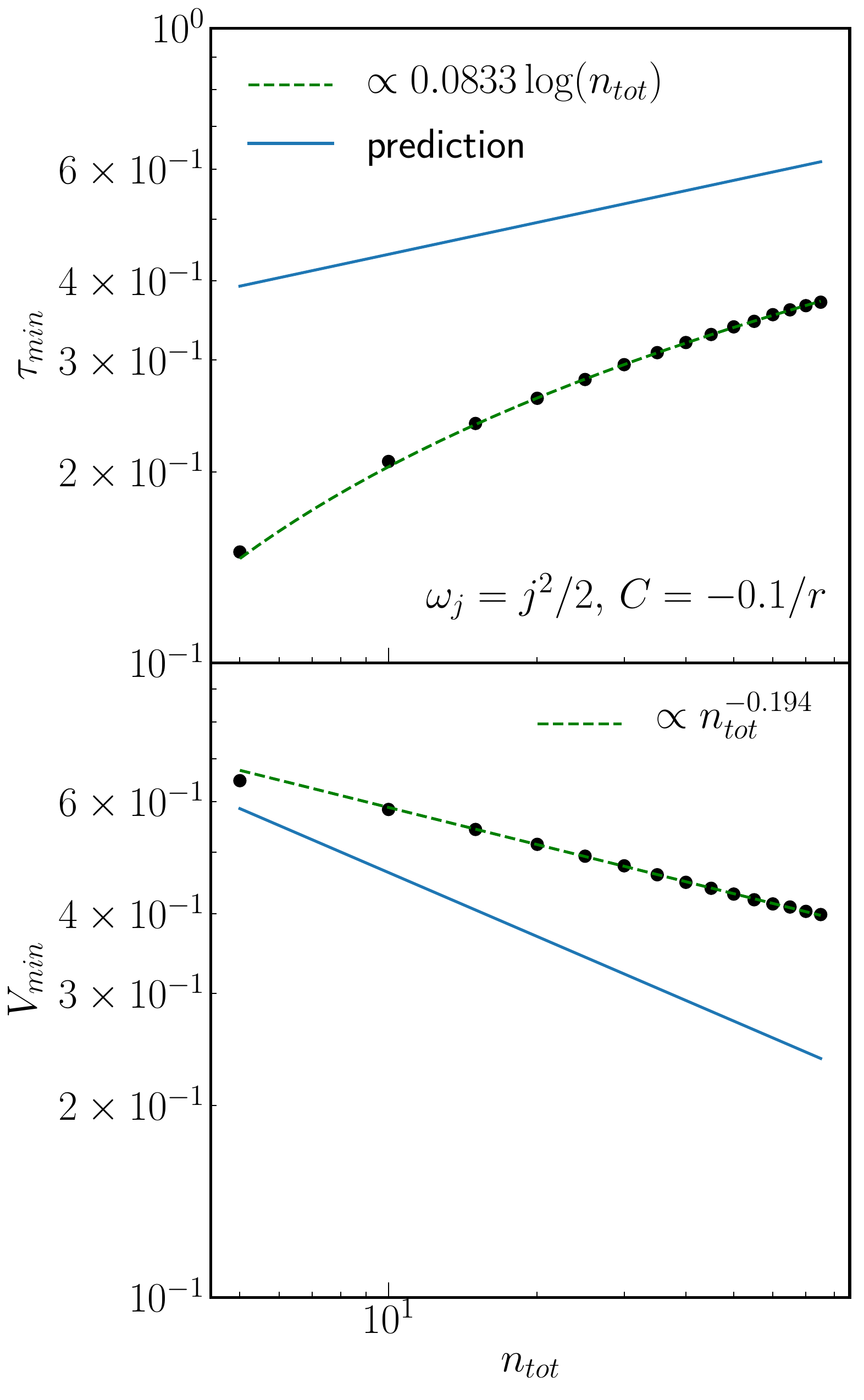}
	\caption{ Here we show the scaling of the minimum value of $V_-$ and the time when it occurs and the best power law fit to the data points. Neither has the same scaling predicted by the one mode model. By the maximum squeezing time the one mode model approximation is no longer accurate. We only include points for which the one mode model was predicted to be accurate, i.e. with $n_{tot} > 20$. }
	\label{fig:sq_scaling}
\end{figure}

We track the squeezing of the operator defined in equation \eqref{a_op}, $V_-(t)$, for systems similar to scalar field dark matter (those with long range interactions, quadratic dispersions, and coherent state initial conditions). The predictions made in \ref{sq_subsection} rely on the validity of the Hartree ansatz. For coherent state initial conditions this assumption is valid at early times. At early times the linear contribution to squeezing from equation 37 in \cite{Eberhardt2021} dominates. This means that squeezing is initially proportional to the classical average particle potential energy, which we hold constant because the dark matter density is well constrained. Therefore, we expect the initial onset of squeezing not to depend on the total occupation number $n_{tot}$, see also the discussion in \cite{KoppFragkosPikovski2021}. We can see in figure \ref{fig:t_sq} that this is the case. Therefore, there is some squeezing that does not go away even as the occupation number is taken to be much larger than unity. Note that for these systems $n_{tot} |\chi| = 0.242$ is constant.

The single mode model also makes a prediction of the maximum squeezing and the time at which this occurs. However, for the multimode simulations run here we see that these do not conform to the predictions assuming the validity of the Hartree ansatz. This is somewhat surprising as the maximum squeezing occurs at a time when the single mode description of the field clearly seems to still be approximately correct. The scaling of the maximum squeezing and maximum squeezing time compared to the Hartree predictions are shown in figure \ref{fig:sq_scaling}.
It can be observed from figure \ref{fig:sq_scaling} that the strength of squeezing, as quantified by $V_{min}$, scales approximately as $n_{tot}^{-0.194}$, implying that highly occupied states would get strongly squeezed at time $\tau_{min}$. 
This time when squeezing becomes minimal is expected to increase with $n_{tot}$ as $n_{tot}^{1/6}$ assuming a Hartree ansatz. Here we observe a milder logarithmic scaling  $0.0833 \log(n_{tot})$, which means that extremal  values of squeezing for large occupation numbers will be reached in just a couple of dynamical time scales. For these fits we use data points which are expected to be well approximated by the Hartree ansatz, i.e. with $n_{tot} \ge 20$. 
In \cite{KoppFragkosPikovski2021} the squeezing time scale was the defined by $V_-$ crossing below $e^{-2}$. Extrapolating from the $V_-$ graph in figure \ref{fig:sq_scaling}, solving $0.919 \,  n_{tot}^{-0.194} = e^{-2}$, we then expect that for $n_{tot} > 1.94\times10^4$ we could define the squeezing time scale $\tau_{sqz, e^{-2}}$ again by $V_-$ crossing below $e^{-2}$. 
We then expect that $\tau_{sqz, e^{-2}} \simeq \tau_{min}(n_{tot}= 1.94\times10^4) = 0.0120 + 0.0833 \log(1.94\times10^4) = 0.834$, which is close to the value $0.6$ found using the Hartree ansatz \cite{KoppFragkosPikovski2021}.
Thus, while the precise scaling of $V_{min}$ and $\tau_{min}$ with $n_{tot}$, the absolute values of $\tau_{sqz, e^{-2}}$, $V_{min}$ and $\tau_{min}$ for fixed $n_{tot}$, were not fully captured by the Hartree ansatz assumed in \cite{KoppFragkosPikovski2021}, the qualitative picture and a priori existence of squeezing of an initially coherent state remains essentially the same.  
Whether squeezing of a dark matter scalar field is observable in principle depends on whether the decoherence time scale is longer than the squeezing time scale, and on whether the pointer states of the dark matter scalar field are coherent states (for  squeezing to be observationally relevant it would not matter if squeezing occured on coherent states if they are not approximately pointer states).
Thus the answer to what extent the squeezing is observationally relevant requires study of the impact of a monitoring and decohering environment, such as baryons, which will be part of future work.

\section{Conclusions} \label{sec:conclusion}

We can see immediately that both total occupation numbers and initial conditions determine the length of time the quantum evolution of a system will track the classical. While number eigenstate, coherent state, and field number state initial conditions can all produce the same mode number occupation expectation values only the latter two approach the results of the MFT as occupation numbers are increased, see figure \ref{fig:qualConv}. This corroborates results found in \cite{Sikivie2017, Hertzberg2016, chakrabarty2021}. 

All of metrics used to measure the quantum breaktime give approximately similar results, over the range of occupation numbers tested, indicating a robustness to the specific definition of breaktime, see figures \ref{fig:params}. The exception is long range system with quadratic dispersion. We can see that, for the system we tested with long range interactions, the breaktime defined using the $Q$ parameter scaled logarithmically with the total occupation number for coherent state initial conditions. This corroborates the arguments made for chaotic systems in \cite{Hertzberg2016,Albrecht2014}. However, the breaktime measured using the PO criterion has power law scaling. Because the $Q$ parameter measures spreading around the mean field value and the PO criterion simply measures whether the $\hat M_{ij}$ matrix can be described by a single field, this corroborates the idea that the density may admit a single field description but that the classical field equations of motion may still admit quantum corrections.

The other systems we tested evolve with Hamiltonians similar to those studied in \cite{Hertzberg2016, Sikivie2017}, with linear dispersion relations and contact interactions. The breaktimes of these systems approximately scales as power laws, roughly $\propto \sqrt{n_{tot}}$. 

We find that the scaling of the breaktime depends on both the initial conditions and interactions. It should be noted that power law and logarithmic scaling make very different predictions for the breaktimes of large systems. It is therefore not clear that studying toy models with only contact interactions can be extrapolated to those with long range interactions. Likewise, it is equally important that initial conditions be appropriately modeled as for systems like our long range toy model all three initial conditions we tested provided different breaktime scaling. 

We find that the onset of quantum squeezing is independent of occupation number and well predicted by the linear approximation at early times, corroborating the results of \cite{KoppFragkosPikovski2021}. This is a result of the fact that the system is initially well described by the Hartree ansatz. The predictions that the single mode model makes for the maximum squeezing value and time however are not accurate for multi mode systems. By the time the state is maximally squeezed it is no longer well approximated by a single mode. 

Coherent state initial conditions and a Hamiltonian which includes long range interactions and a quadratic dispersion represents the system most similar to scalar field dark matter. Because our results indicate a logarithmic enhancement of the breaktime with particle number, it is not immediately clear that quantum corrections would remain small over the age of the universe for cosmologically interesting systems even for ultralight dark matter candidates. Therefore, work investigating the breaktime for more realistic systems is needed to conclude whether or not quantum corrections are relevant to the evolution of scalar field dark matter. An analysis of this question, using the solver introduced in \cite{Eberhardt2021}, will be included in subsequent work. 

\begin{acknowledgments} 
We would like to thank Mark Hertzberg, Erik Lentz, Anthony Mirasola, Chanda Prescod-Weinstein, and Pierre Sikivie for helpful comments and discussion. Some of the computing for this project was performed on the Sherlock cluster. A.E., A.Z., and T.A. are supported by the U.S. Department of Energy under contract number DE-AC02-76SF00515. 
\end{acknowledgments}

\appendix*

\section{Special Hilbert space size} \label{app:spHS}

\begin{figure}
	\includegraphics[width = .45\textwidth]{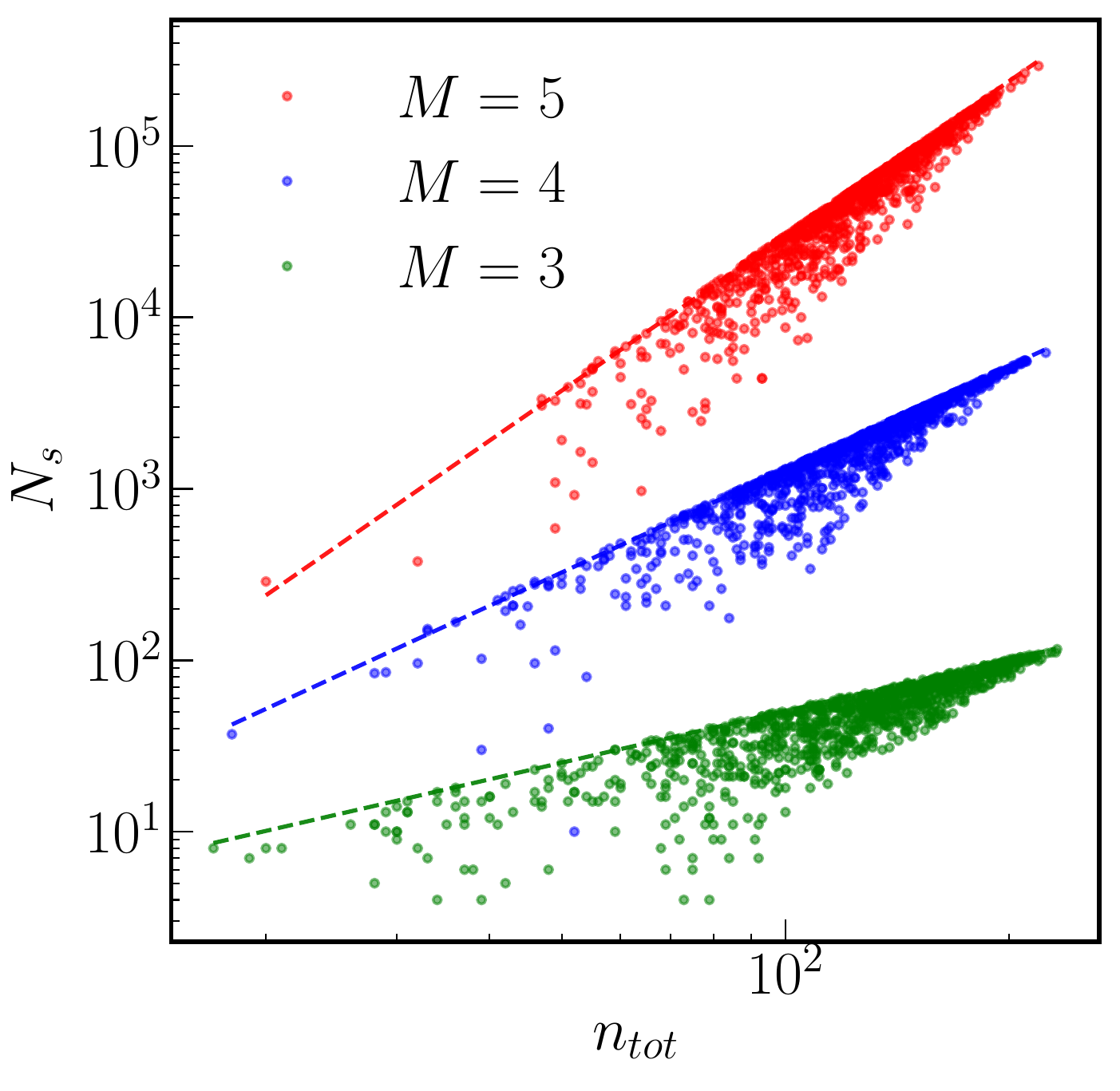}
	\caption{ Here we show the size of special Hilbert spaces with random $n_{tot}$ and $p_{net}$ as a function of $n_{tot}$ for a number of different $M$. The solid lines are $\propto n_{tot}^{M-2}$ shown for reference. We can see that the special Hilbert spaces have this approximate scaling. }
	\label{fig:scaling}
\end{figure}

In this appendix we discuss how the size of the special Hilbert spaces scales with the simulation parameters. Let us consider a system of $M$ interacting modes. Recall that a given special Hilbert space is uniquely identified by a net momentum and total particle number as 

\begin{align}
    n_{tot} &= \sum_{n_k \in \set{n}} n_k \, \textrm{, and} \\
    p_{net} &= \sum_{n_k \in \set{n}} k \, n_k \, .
\end{align}

And that the space contains all the number eigenstates which have the same $n_{tot}$ and $p_{net}$. The dimensionality of the special Hilbert space then is simple the number of ways it is possible to place $n_{tot}$ in $M$ modes such that the net momentum is $p_{net}$. The simulations we run here typically approximately satisfy the conditions that $n_{tot} \gg M > 1$. In this limit the possible arrangements go as $\binom{n_{tot}}{M-2} \sim n_{tot}^{M-2}$, the $-2$ coming from the two constraints, fixed net momentum and total particle number. 

To demonstrate this scaling we plot the size of special Hilbert spaces containing a number eigenstate with a random number of particles placed in each mode. This is shown in Figure \ref{fig:scaling}. We can see that the scaling approximately follows 
$\sim n_{tot}^{M-2}$.

It should also be noted that each individual Hilbert space requires far fewer numbers to represent than predicted by equation 5 in \cite{Sikivie2017}.

\section{Relaxation time}

\begin{figure}
	\includegraphics[width = .45\textwidth]{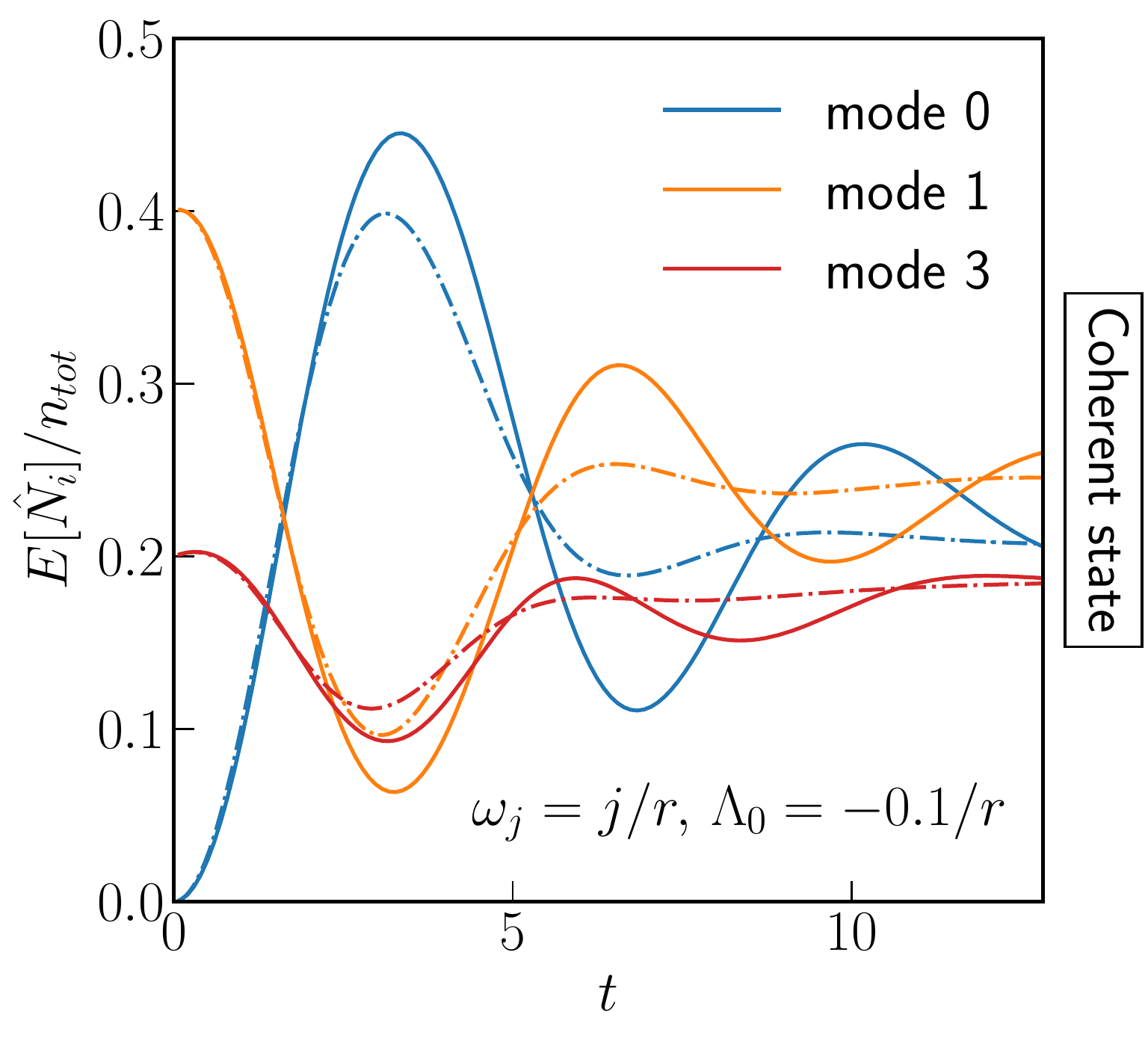}
	\caption{ Here we plot the evolution of three mode occupation numbers for coherent state initial conditions with attractive contact interactions. We show the results of two different simulations, $r=15$ is shown in the solid lines and $r=5$ in dash dotted lines. We can see that the oscillations in the occupation numbers increase in amplitude and duration with occupation number. This implies that the relaxation time increases with occupation number, and we can see that at times when the smaller total particle number simulation is relaxed the larger is still dynamically evolving. This is consistent with coherent state initial conditions approaching the classical evolution as occupation number is increased.}
	\label{fig:csRelax}
\end{figure}

\begin{figure}
	\includegraphics[width = .45\textwidth]{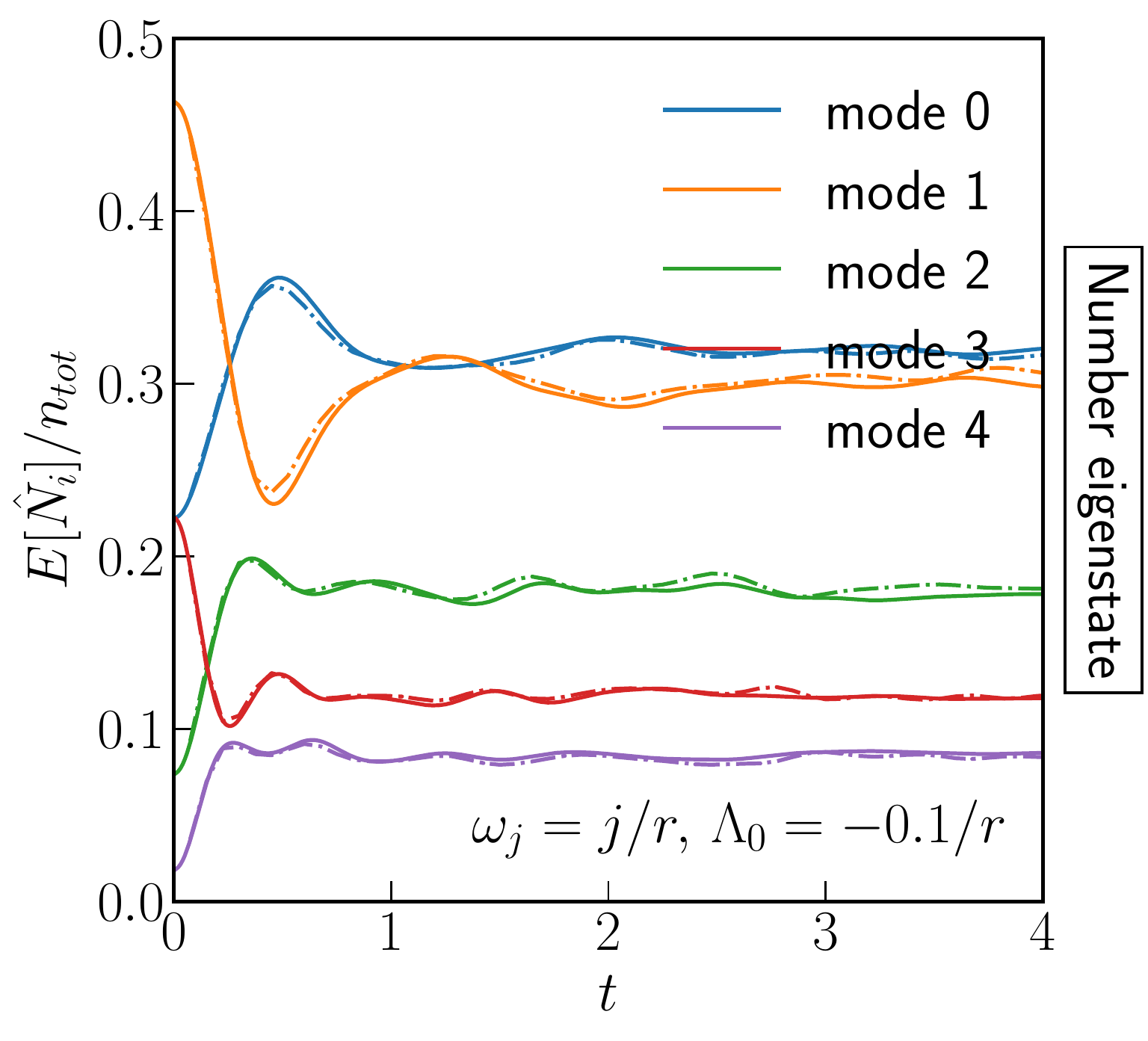}
	\caption{ Here we plot the evolution of mode occupation numbers for number eigenstate initial conditions with attractive contact interactions, note that this is the same system and initial conditions studied in \cite{Sikivie2017}. We show the results of two different simulations, $n_{tot}=162$ is shown in the solid lines and $r=54$ in dash dotted lines. We can see that the evolution is similar between both simulations. This implies that the relaxation time is fixed for these systems at high occupation. This is consistent with number eigenstate initial conditions not approaching a single classical field evolution as occupation number is increased.}
	\label{fig:neRelax}
\end{figure}

For number eigenstate initial conditions it was demonstrated in \cite{Sikivie2017} that there was a fixed relaxation time that depended only on constants present in the classical theory as the total occupation number of states became large, see figure \ref{fig:neRelax}. Likewise, it was shown that the classical evolution never relaxed. We would therefore expect that if coherent states approached the classical solution with increasing occupation number that the relaxation time of our quantum simulations would also increase with occupation number. This is exactly what we see for coherent state initial conditions in figure \ref{fig:csRelax}, higher occupation number states oscillate more dramatically and have a longer relaxation time, consistent with them approaching the classical field description in the high occupation number limit.

\bibliography{BIB}

\end{document}